# A Faint Luminous Halo that May Trace the Dark Matter around Spiral Galaxy NGC 5907

Penny D. Sackett[1]
Institute for Advanced Study, Princeton, NJ 08540

Heather L. Morrison[2]
National Optical Astronomy Observatories, Tucson, AZ 85726

Paul Harding
Steward Observatory, University of Arizona, Tucson, AZ 85721

Todd A. Boroson
U.S. Gemini Project Office, National Optical Astronomy Observatories, Tucson, AZ 85726

**The presence of unseen halos of "dark matter" has long been inferred from the high rotation speeds of gas and stars in the outer parts of spiral galaxies[1]. The volume density of this dark matter decreases less quickly from the galactic center than does that of the luminous mass (such as that in stars), meaning that the dark matter dominates the mass far from the center[1,2]. While searching for faint starlight away from the plane of the edge-on disk galaxy NGC 5907[3], we have found that the galaxy is surrounded by a faint luminous halo. The intensity of light from this halo falls less steeply than any known luminous component of spiral galaxies, but is consistent with the distribution of dark mass inferred from the galaxy's rotation curve.**

---

[1]J. Seward Johnson Fellow

[2]Hubble Fellow

In order to characterize faint extended stellar emission around galaxies it is necessary to (1) calibrate precisely and accurately the detector response, (2) obtain accurate measurements of the sky brightness at distances that avoid contamination from galaxy light, and (3) have many spatial resolution elements across the galaxy. These requirements motivate the use of a large format CCD on a wide field telescope to study large, edge-on systems. We have carried out extremely deep photometry of the thin, edge-on (i = 87°) Sc galaxy NGC 5907 using a 2048 × 2048 CCD on the KPNO 0.9m telescope over a field size of about 24′ square. By obtaining 6 1800 second galaxy exposures, in which the image was displaced ~30 pixels (corresponding to ~3 disk scale heights), and 15 dark sky flats of 1800 seconds exposure each, we have been able to characterize the response of the detector (flat-field) to 0.02% of sky in 10×10 pixel (8″×8″) bins, and to 0.1% over the large scales (~100″ or ~5 kpc) typical of the extended halo light. (This high level of flat-fielding accuracy has been achieved in previous work on faint galaxies[4,5].) The sky level of R=20.43 mag arcsec$^{-2}$ has been determined to 0.01% from the edges of the galaxy frames and subtracted. An accurate error model including the effects of photon noise, flat-fielding errors, readout noise, and surface brightness fluctuations allows us to reach R=27 mag arcsec$^{-2}$ reliably. Full details concerning the data acquisition, reduction, and error analysis are reported elsewhere[3].

A spiral galaxy is composed of a thin disk of young stars (called Population I stars) whose local surface brightness falls exponentially with cylindrical distance from the galactic center and with height above the galactic plane. In addition, one or more of the following stellar components is generally present[6]: (1) a thick disk with larger scale height, (2) a centrally condensed bulge confined to the inner regions of the galaxy, (3) an extended spheroidal stellar halo composed of old (Population II) stars found as field stars or in globular clusters. (The precise relationship between bulges and Pop II halos remains unclear.) In addition to differences in kinematics, these components are distinguished by differences in spatial scales and surface brightnesses: the thin disk and bulge are brighter and less extended than the thick disk and stellar halo. Known Pop II field stars and globular cluster systems have luminosity volume densities that fall steeply with distance ($r$) from the galactic center: Milky Way globular clusters have an $r^{-3.5}$ density distribution[7], while the M31 halo field star distribution is $r^{-4}$ or steeper[8]; similar results are seen for NGC 4565[9].

An original goal of this program was to search for a thick disk in NGC 5907. No thick disk was detected, but extended light (in excess of the thin disk) was seen below R≈25 mag arcsec$^{-2}$ in all profiles parallel to the minor axis at distances corresponding to 7–16 thin disk scale heights (3–7 kpc) above the galactic plane (Fig. 1). Since the surface brightness of this extended light is typically 1% of sky over these regions, neither an inappropriate choice of sky level nor flat-fielding errors can account for its presence. Furthermore, the faint halo is not due to scattered light: by using bright, uncrowded stars (some saturated) to derive a point spread function out to 3′ (10 kpc), we have deconvolved the image and see no change in the extended light[3]. The extremely small bulge observed in the infrared[10] is too centrally concentrated to account for the excess light, and detailed modeling has indicated that the extended light is not due to a thick disk[3], since a scale height ∼3 times that of any known thick disk (and a normalization at least an order of magnitude less) would be needed.

We therefore model the halo light of NGC 5907 with a flattened, power law distribution with a core, applying a non-linear least squares routine to fit thin disk + power law halo models to the total light. (Inside the core, the model volume density is roughly constant; outside the core the density drops as an inverse power of the radius.) Our least-squares routine is capable of determining the power-law exponent to within 0.20 and the density axis ratio $c/a$ (*i.e.,* the flattening) to within 0.15; these estimates were obtained in experiments using artificial data, in which the power law behavior of the profile is known, in the presence of realistic noise. The resulting weighted residuals after the model has been subtracted from the photometric data of NGC 5907 are shown in Fig. 2. When the halo exponent is allowed to float as a free parameter, the best fits to the extended light with small cores ($<$ 1 kpc, characteristic of known stellar halos) have an exponent of about $-2.2$ and an axis ratio of about 0.5; models with a larger core radius of 6 kpc have an exponent of about $-2.7$ and similar axis ratio.

The surface brightness that we observe for the extended light is appropriate for a stellar halo of Pop II field stars[11], but its radial profile is much shallower than the $r^{-3.5}$ profile typical of Pop II field stars and globulars. (NGC 5907 has been searched for globular clusters[12] but none were detected.) To our knowledge, this is the first detection of an extended luminous component measured to have an approximately $r^{-2}$ profile around a spiral galaxy. Only in the outer regions of cD galaxies, which have a totally different environment, has a similarly shallow radial profile been observed[13]. The shallowness of its profile suggests that the faint halo of NGC 5907 has suffered little from the dynamical evolution processes such as dissipation and relaxation that produce the central concentration and steep profiles of typical stellar halos[14].

Since the faint extended light of NGC 5907 is not well fit by the steep profile of a typical Pop II halo (Figs. 1, 2), we are led to consider how it may be related to the only component of spirals known to have a shallow distribution: the unseen "dark halos" that are inferred to have intrinsic $r^{-2}$ *mass* profiles. (Somewhat steeper profiles have also been suggested[15].) Dark halos are generally thought to have core radii of 1–10 kpc[16,17] and to be significantly flattened[18], but the nature of their constituents remains unknown. The speculative hypothesis that the faint light that we have detected traces (with constant mass-to-light ratio $M/L_R$) the dark matter halo of NGC 5907 predicts that a simple rescaling of the halo light would produce, in combination with the other luminous components of the galaxy, a model capable of reproducing the observed kinematics. In Fig. 3 we show that a remarkably good match to the rotation curve of NGC 5907[2] can be obtained using a mass model containing only a (stellar + gasous) disk and a massive halo with shape and scale parameters derived from the fits to the observed light extrapolated over the radial range of the kinematical data.

Could the faint halo light of NGC 5907 be composed of dim, low-mass stars that not only trace the dark matter, but are also responsible for all of its dark mass? By varying only two parameters in the rotation curve fit, the $M/L_R$ of the disk and halo, it is possible to place constraints on the density normalization of the halo required to achieve a good fit; these constraints require that $270 \lesssim M/L_{R,halo} \lesssim 540$ in order to explain simultaneously the faint extended light around NGC 5907 and its dynamics. Recent reports of microlensing in the Milky Way[19,20,21] suggest that a measurable fraction of the dark matter in our own Galaxy may be in the form of compact objects of roughly one-tenth the mass of the Sun, near the hydrogen burning limit for stars. The magnitudes and colors of M dwarfs and brown dwarfs, which have masses in this range, are strong functions of their metallicity[22,23,24]. If such objects make up the dark halo, they would be expected to have heavy element abundances at least as small as those of Pop II stars. Such stars are seen in the solar neighborhood, namely the late M subdwarfs of the Galaxy's Pop II halo[22]. The average $M/L_R \approx$450 that we derive for the putative "dark" constituents of NGC 5907 is slightly lower than the $M/L_R \approx 600$ of faint observed late M subdwarfs[24], but higher than the $M/L_R \approx 90$ derived from models of zero metallicity M dwarfs near the hydrogen-burning limit[23]. Detection limits are strongly dependent on wavelength; non-detections of extended light in NGC 5907 from older photometry in the B[25] and K[26,10] bands are consistent with the colors of observed late M-dwarfs. Models of zero-metallicity brown dwarfs give $M/L_R \approx 30{,}000$[24]; brown dwarfs are thus too dim to account for the halo light of NGC 5907.

Since halo stars have high space motions, proper motion surveys provide one of the best constraints on the existence of a massive halo of M subdwarfs in our own Milky Way. Assuming kinematics similar to the Galaxy's Pop II halo, the numbers of M subdwarfs in one such proper-motion catalog[27] fall short by at least an order of magnitude[28], suggesting that the Galaxy's massive halo is not composed of such objects. Deep pencil beam surveys of the Milky Way capable of accurate star-galaxy separation at V=26 or fainter would also place tight constraints on a dark matter halo of late M subdwarfs in the Galaxy since they would probe distances up to 2 kpc and thus detect ≈20 of these faint halo stars per square arcminute.

For external galaxies, the decisive observations, which are in reach of current technology, will be deep V and I surface brightness measurements. If the stellar population emitting the faint halo light of NGC 5907 is similar to that of known Pop II stars, its V−I color should be about 1.0[29]; observed late M-dwarfs have V−I colors near 2.5[22].

We are grateful to Renzo Sancisi for supplying the HI rotation curve of NGC 5907 and to Adam Burrows and Jim Liebert for illuminating discussions about low-mass stars. We thank Hans-Walter Rix for acting as scientific matchmaker and Ken Freeman and Leonard Searle for suggestions that improved the presentation of this paper.

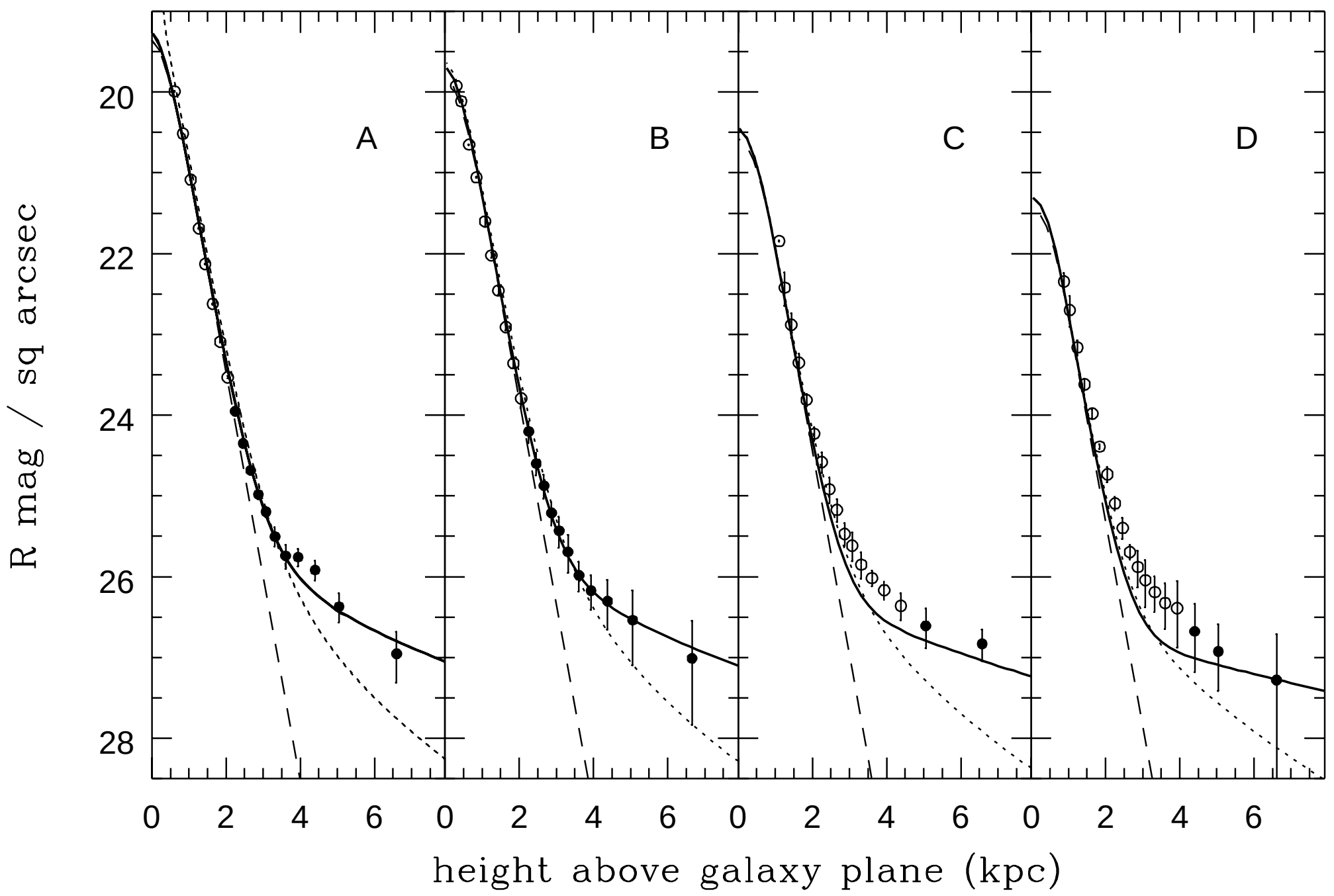

Fig. 1.— The total surface brightness in R mag arcsec$^{-2}$ is shown as a function of distance from the galactic plane of NGC 5907 for four different cuts (spaced 4.1 kpc apart) perpendicular to the galactic plane. Positions of the cuts are shown in Figure 2(a). For each profile, light from matching positions in the four quadrants is averaged. Error bars include differences between quadrants as well as individual errors. Distances are in kpc and assume a distance of 11 Mpc to the galaxy. Open symbols denote areas underneath the mask shown in Figure 2, which do not contribute to the halo fit; closed symbols indicate unmasked areas. The dashed line shows the best fit to the total projected light using a thin, exponential disk, with scale length 4.8 kpc, scale height 430 pc, central surface brightness of 19.2 R mag arcsec$^{-2}$ and truncation radius of 19.5 kpc. To obtain the thin disk fit, the inner regions of the galaxy were masked to avoid contamination from dust[3]. Also shown are the best fit for disk plus $r^{-3.5}$ halo with a 200 pc core (dotted line), and disk plus $r^{-2.26}$ halo with a 2 kpc core (solid line).

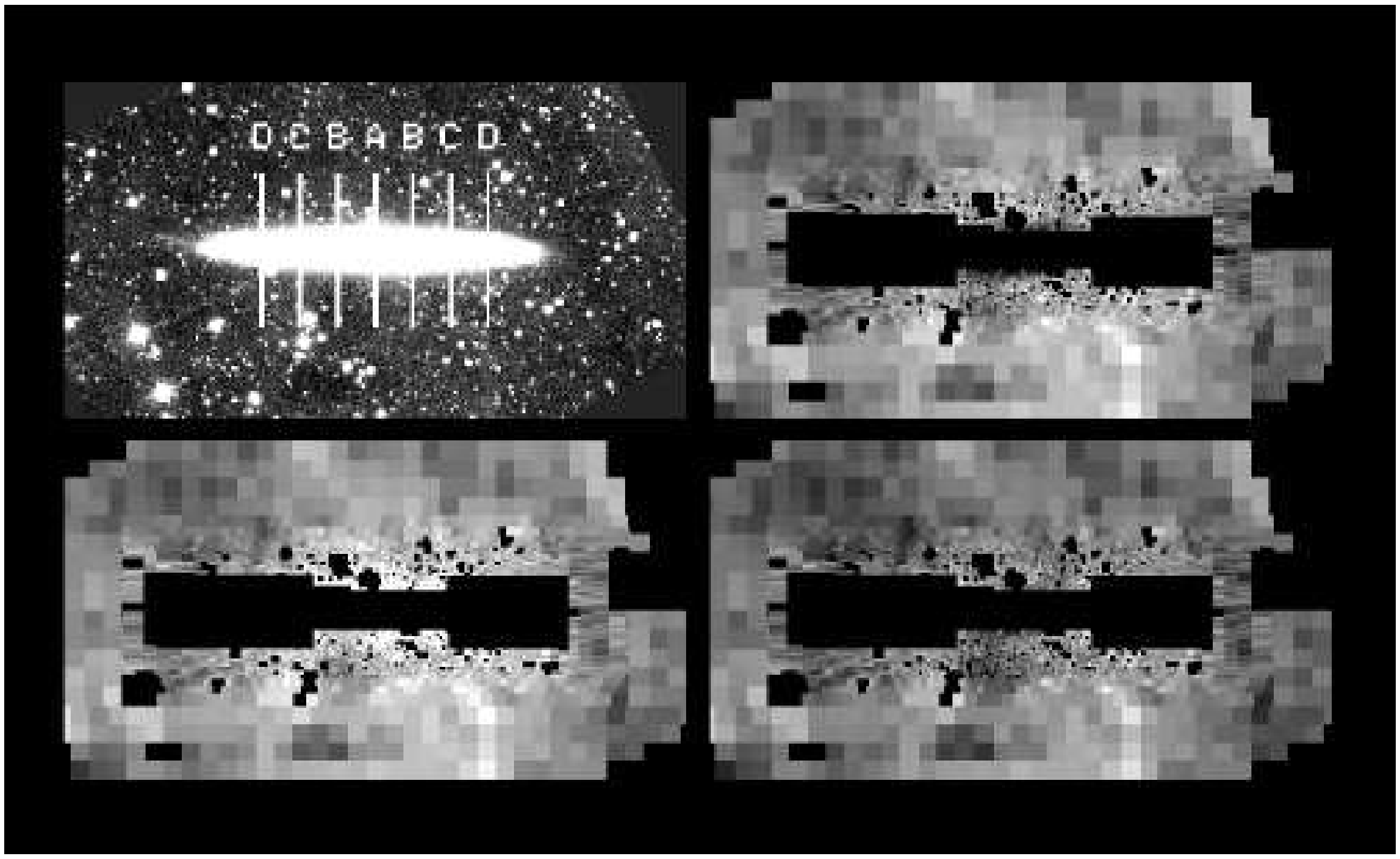

Fig. 2.— Optical image of NGC 5907 (top left) and weighted residual images for three models. The area shown in each frame is slightly less than one-half the total usable detector area. Light areas indicate regions of excess light compared to the given model; dark areas indicate regions of insufficient light; black shows regions that have been masked due to confusion from dust, foreground stars, and warp. The data have been binned into $4'' \times 4''$ bins near the galaxy and $40'' \times 40''$ bins in the halo and sky regions in order to achieve similar signal-to-noise in each bin; thus each bin, regardless of its size, receives a similar weight in the fit. The residuals are expressed in terms of the uncertainty in each bin: white indicates regions with light in excess of $4\sigma$ from the model and black indicates $4\sigma$ deficits of light. The masked region is 46 kpc long and 4.3 kpc wide at the minor axis. **(top left)** Greyscale R-band image with the placement of the vertical profiles shown in Fig. 1 superposed. The contrast has been adjusted so that the galaxy is "over-exposed"; essentially *all* light from the exponential disk is shown as pure white. **(bottom left)** Residuals from Thin Disk Only model. Disk parameters are given in the caption to Fig. 1. The strong positive residuals indicate the presence of extended halo light. Reduced $\chi^2$ for this fit is 6.0. **(top right)** Residuals from Thin Disk + $r^{-3.5}$ Halo model. Disk parameters are given in the caption to Fig. 1; halo has $c/a = 0.27$ and a core radius of 200 pc. Reduced $\chi^2$ for this fit is 2.3. **(bottom right)** Residuals from Thin Disk + Free Power Law Halo model. Disk parameters are unchanged; halo has an exponent of $-2.26$, $c/a = 0.53$, and a core radius of 2 kpc. Reduced $\chi^2$ for this fit is 1.5.

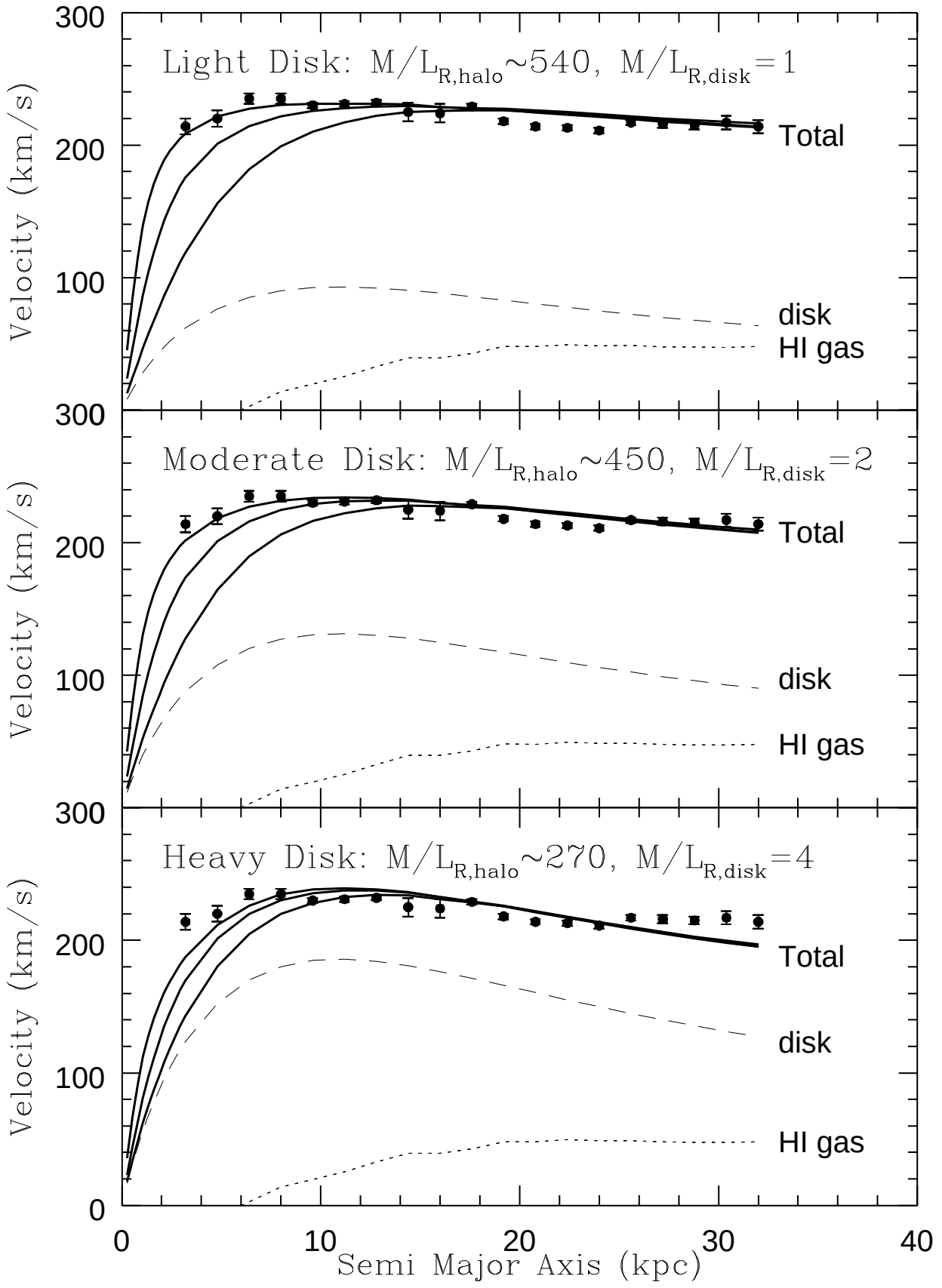

Fig. 3.— The rotation curve of the galaxy as derived from 21cm line observations of neutral hydrogen[2] is shown with model curves. The dotted line shows the curve that would be expected from the HI gas alone and the dashed line indicates the rotation curve of the thin exponential disk alone. The thick solid lines are the complete rotation curves, namely the quadrature sum of the gas curve, disk curve, and the rotation curve of halos producing similar good fits to the light: an $r^{-2.22}$ halo, with $c/a = 0.51$ and a 1 kpc core (top curve), an $r^{-2.26}$ halo, with $c/a = 0.53$ and a 2 kpc core (middle curve; R-band residuals shown in Fig. 2d), and an $r^{-2.74}$ halo, with $c/a = 0.53$ and a 6 kpc core (bottom curve). The central dark mass densities of these models are 1.9, 0.45 and 0.08 $M_{\odot}/\mathrm{pc}^3$ respectively. In the middle panel, the mass of the disk is assumed to be $5.3 \times 10^{10}$ $M_{\odot}$, derived using an $M/L_R = 2$ (appropriate to Sc galaxies) and a luminosity based on the Tully-Fisher relation. The $M/L_R$ of the halo component has been scaled to give a good fit to the rotation curve: $M/L_R = 420$ for the $r^{-2.22}$ halo, $M/L_R = 450$ for the $r^{-2.26}$ halo, and $M/L_R = 455$ for the $r^{-2.74}$ halo. Plausible fits to the curve can be obtained with plausible $1 \lesssim M/L_{R,disk} \lesssim 4$ (top, middle, and bottom panels) for an Sc disk, resulting in $270 \lesssim M/L_{R,halo} \lesssim 540$ for the halo, though the high $M/L_{R,disk}$ and low $M/L_{R,halo}$ give notably poorer fits. (All $M/L$ are given in solar units.) The $r^{-3.5}$ distribution cannot provide the required rotation support beyond the edge of the stellar disk ($\approx$20 kpc). The rotation curve inward of 5 kpc is affected by light inside our mask and by the small bulge component seen in the infrared[10].